\documentclass[
 aps,
 11pt,
 final,
 notitlepage,
 oneside,
 nobibnotes,
 nofootinbib,
 superscriptaddress,
 noshowpacs,
 centertags]
{revtex4}

\def\squareforqed{\hbox{\rlap{$\sqcap$}$\sqcup$}}

\def\sq{\ifmmode\squareforqed\else{\unskip\nobreak\hfil
\penalty50\hskip1em\null\nobreak\hfil\squareforqed
\parfillskip=0pt\finalhyphendemerits=0\endgraf}\fi}

\def\degr{\hbox{$^\circ$}}

\def\arcmin{\hbox{$^\prime$}}

\def\arcsec{\hbox{$^{\prime\prime}$}}

\def\utw{\smash{\rlap{\lower5pt\hbox{$\sim$}}}}

\def\udtw{\smash{\rlap{\lower6pt\hbox{$\approx$}}}}

\def\fs{\hbox{$\,.\!\!^{\rm s}$}}

\def\diameter{{\ifmmode\mathchoice
{\ooalign{\hfil\hbox{$\displaystyle/$}\hfil\crcr
{\hbox{$\displaystyle\mathchar"20D$}}}}
{\ooalign{\hfil\hbox{$\textstyle/$}\hfil\crcr
{\hbox{$\textstyle\mathchar"20D$}}}}
{\ooalign{\hfil\hbox{$\scriptstyle/$}\hfil\crcr
{\hbox{$\scriptstyle\mathchar"20D$}}}}
{\ooalign{\hfil\hbox{$\scriptscriptstyle/$}\hfil\crcr
{\hbox{$\scriptscriptstyle\mathchar"20D$}}}}
\else{\ooalign{\hfil/\hfil\crcr\mathhexbox20D}}%
\fi}}

\usepackage[utf8]{inputenc}
\usepackage[russian,english]{babel}
\usepackage{graphicx}

\usepackage{float}
\usepackage{amsmath}
\usepackage{multirow}

\begin{document}
\selectlanguage{english}


\keywords{methods: observational--techniques: photometry--galaxies: active}

%


\title{Photometric reverberation mapping of AGNs at $0.1 < z < 0.8$: I.  \\ Observational technique}


\author{\firstname{R.~I.}~\surname{Uklein}}
 \email{uklein@sao.ru}
 \affiliation{Special Astrophysical Observatory of the Russian Academy of Science, Nizhnii Arkhyz, 369167, Russia}


\author{\firstname{E.~A.}~\surname{Malygin}}
 \affiliation{Kazan Federal University, Kazan, 420008 Russia}


\author{\firstname{E.~S.}~\surname{Shablovinskaya}}
\affiliation{Special Astrophysical Observatory of the Russian Academy of Science, Nizhnii Arkhyz, 369167, Russia}


\author{\firstname{A.~E.}~\surname{Perepelitsyn}}
\affiliation{Special Astrophysical Observatory of the Russian Academy of Science, Nizhnii Arkhyz, 369167, Russia}


\author{\firstname{A.~A.}~\surname{Grokhovskaya}}
\affiliation{Special Astrophysical Observatory of the Russian Academy of Science, Nizhnii Arkhyz, 369167, Russia}


\begin{abstract}
The improvement of the calibration relation for determining the
size of the broad-line region from the observed optical luminosity
of active galactic nuclei (AGN) is a necessary task to study
fundamental parameters of distant AGNs such as mass of the central
supermassive black hole. The most popular method of the BLR size
estimation is the reverberation mapping based on measuring the
time delay between the continuum flux and the flux in the emission
lines. In our work, we apply the method of photometric
reverberation mapping in medium-band filters, adapted for
observations at the Zeiss-1000 telescope of the SAO RAS, for the
study of AGN with broad lines in the range of redshifts $0.1 < z <
0.8$. This paper describes the technique of observations and data
processing, provides a sample of objects and demonstrates the
stability of the used method. As a preliminary result for
2MASX\,J08535955+7700543 at $z=0.1$ we have obtained time delay
estimates of $\tau (ICCF)=32.2^{\pm10.6}$ days and $\tau({\rm
	JAVELIN})=39.5_{-15.8}^{+23.0}$ days that are consistent with each
other and also within the accuracy of the existing calibration
relations.
\end{abstract}

\maketitle

\begin{center}
Received: December 26, 2018. \quad
Accepted: September 23, 2019.
\end{center}

\section{Introduction}
\label{S:int}

According to modern concepts, in the centers of most massive
galaxies, there are supermassive black holes (SMBH) weighing from
a million to tens of billions of solar masses. In the active
galactic nuclei (AGN), the accretion of gas onto a supermassive
black hole leads to the release of a huge amount of energy.
Variable accretion disc radiation in the ultraviolet part of the
spectrum ionizes gas clouds in the broad-line region (BLR), which
then re-emit energy in the emission lines due to
photorecombination. The reverberation mapping method (RM) consists
in measuring the time delay $\tau$ between the radiation of the
accretion disk responsible for the formation of the continuum and
the radiation in the emission lines produced in the BLR region
\cite{BlanfordMKee82}. It is assumed that the size of
the BLR $R_{\rm BLR} \equiv c\tau$, where $c$ is the speed of
light \cite{Peterson93}. Then $R_{\rm BLR}$ can be
determined by measuring the time lag $\tau$ of the light curve in
an emission line relative to the one in the continuum associated
by an integral transformation with the cross-correlation function
kernel.

The first works on the measurement of the time delay $\tau$
between  H$_\alpha$ and the ultraviolet continuum radiation in
galaxies with active nuclei were carried out in the paper by
Cherepashchuk and Lyutyi in 1973 \cite{CherepLyut73}.
For the investigating galaxies NGC\,4151, NGC\,3516 and NGC\,1068
the delays were found---30, 25 and 15~days, respectively, which
gives the BLR size $c\tau$ $\sim$ 0.02~pc. Reverberation mapping
method was developed in several papers (e.g.,
\cite{BlanfordMKee82, AntBoch83,
	GaskellSparke86} and others).

The data on the BLR size gained for AGNs by RM during the previous
15 years were collected and analyzed in the paper by Peterson et
al. \cite{Peterson04}. The paper by Bentz and Katz 2015
\cite{BentzKatz15} describes a database with a compilation of all
published at that time 63 AGNs with the estimation of central
black holes masses. However, the database \cite{BentzKatz15}
contains observational data only for the nearest AGNs up to
$z\sim0.3$ as well as many later RM investigations (e.g.
\cite{Du16,Jiang16,Nunez17}).

RM of distant AGNs is particularly interesting. Since the gas
dynamics in the BLR is influenced by the SMBH gravitation,
according to the virial ratio its mass is related to the size
$c\tau$ and the gas velocity in the BLR $\upsilon_{\rm line}$ as:
$$M_{\rm SMBH}=fc\tau\upsilon_{\rm line}^2G^{-1}, $$
where $G$ is the gravitational constant, $c$ is the speed of
light, $f$ is a dimensionless factor of the order of one depending
on the BLR structure and kinematics and the angle of the system
inclination relative to the observer.

Thus, the extension of the AGN sample with known sizes $R_{\rm
	BLR}$ to more distant redshifts will allow one to trace the
evolution of the SMBH masses. The largest campaign of BLR RM at
bigger $z$ is the SDSS-RM project \cite{Shen15} monitoring 849
quasars with broad lines in the range of the redshifts $0.1 < z <
4.5$. SDSS-RM collaboration presents the first results of
measurements of the time delays for 44 and 18 quasars using
H$_\beta$ and H$_\alpha$ lines, respectively, in the $0.11 < z <
1.13$ range \cite{Grier17,Grier18err}. It is also noted that for
distant objects, it is needed to observe the delay $\tau$ in the
lines with higher ionization potentials that are in a shorter
wavelength part of the spectrum compared to the Balmer series. For
these purposes, such lines as Mg\,II could be used (2798~\AA)
(\cite{Czerny19} and references therein).

However, the RM method requires the accumulation of a long series
of observational data making harder its wide application. It has
been observed that for active nuclei there is a relation between
the BLR size and AGN luminosity: $R_{\rm BLR} \propto L^{\alpha}$.
Currently, several empirical relations are linking the size
$R_{\rm BLR}$ obtained by measuring the delay in different lines
and luminosity in different spectral bands. The most popular
relation used for nearest objects is $R_{\rm BLR}-L_{\lambda}$
(5100~\AA), where $R_{\rm BLR}$ is the BLR size measured by the
radiation delay in the line H$_\beta$, and $L_{\lambda}$
(5100~\AA) is the luminosity of the AGN in the range
4400--5850~\AA~\cite{Kaspi05,bentz09}.

Our study is dedicated to complement the existing relation of
$R_{\rm BLR}(L)$ by new measurements of $R_{\rm BLR}$ = $c\tau$
for the distant AGNs up to $z \sim 0.8$ using a sample of objects
that do not overlap with other surveys. Besides, we adopt the
photometric RM method~\cite{Haas11} for mid-band
observation with the 1\mbox{-}meter class telescope (Zeiss-1000
SAO RAS).

In this paper, we describe the observational technique of the
photometric RM monitoring of BLR in AGNs, including the
description of instruments and data processing, a sample of
observed objects with expected time delays estimated from the
literature spectral data as well as methodological results
confirming the stability of the implemented method.

\section{Sample}
\label{S:sam}

To conduct a photometric RM monitoring of BLR a sample of 8 active
nuclei with broad lines (equivalent width $W_{\lambda} \geq
200$~\AA) in the range the redshifts $0.1 < z < 0.8$ was composed
by using the databases NED\footnote{NASA NED
	https://ned.ipac.caltech.edu} and
SDSS\footnote{https://dr14.sdss.org/}. For the observations, the
1-m telescope Zeiss-1000 is involved, and the limit on the
brightness of the object is $m < 20^{\rm m}$. The sample includes
only near-polar objects (Dec $ > 68^{\circ}$) to observe them
throughout the year. The final sample is shown in Table
\ref{T:sample}. Columns are following: (\#) identification number
in the sample; (1) galaxy name; (2) coordinates for the J2000
epoch; (3) magnitude in the $V$ filter; (4) redshift $z$; (5)
observed broad emission line; (6) expected delay $\tau$ in days;
(7) used SED filters to measure fluxes in the line and continuum.

\begin{table*}[]
	\caption{Observed sample of active nuclei}
	\medskip
	\begin{tabular}{c|l|c|c|c|c|c|c}
		\hline
		\# & \multicolumn{1}{c|}{Name} & RA Dec (J2000) & Mag ($V$) & $z$ & Emission & $\tau$, days  & Filters  \\
		\hline
		(1) & \multicolumn{1}{c|}{(2)} & (3)  & (4) & (5) & (6) & (7) & (8) \\
		\hline
		1 & 2MASX\,J08535955+7700543     & $08^{\rm h}53^{\rm m}59\fs4$ $+77\degr00\arcmin55\arcsec$ & $17.0$ & $0.106$ & H${\alpha}$ & $27$  & $725/700$ \\
		2 & VII Zw 244                  & $08^{\rm h}44^{\rm m}45\fs3$ $+76\degr53\arcmin09\arcsec$ & $15.7$ & $0.131$ & H${\beta}$  & $34$  & $550/525$ \\
		3 & SDSS\,J093702.85+682408.3    & $09^{\rm h}37^{\rm m}02\fs9$ $+68\degr24\arcmin08\arcsec$ & $18.0$ & $0.294$ & H${\beta}$  & $47$  & $625/600$ \\
		4 & SDSS\,J094053.77+681550.3    & $09^{\rm h}40^{\rm m}53\fs8$ $+68\degr15\arcmin50\arcsec$ & $19.4$ & $0.371$ & H${\alpha}$ & $59$  & $900/875$ \\
		5 & SDSS\,J100057.50+684231.0    & $10^{\rm h}00^{\rm m}57\fs5$ $+68\degr42\arcmin31\arcsec$ & $19.0$ & $0.499$ & H${\beta}$  & $80$  & $725/700$ \\
		6 & 2MASS\,J01373678+8524106     & $01^{\rm h}37^{\rm m}36\fs7$ $+85\degr24\arcmin11\arcsec$ & $16.6$ & $0.499$ & H${\beta}$  & $79$  & $725/700$ \\
		7 & SDSS\,J095814.46+684704.8    & $09^{\rm h}58^{\rm m}14\fs4$ $+68\degr47\arcmin05\arcsec$ & $19.7$ & $0.662$ & H${\beta}$  & $92$  & $800/775$ \\
		8 & GALEX\,2486024515200490156   & $10^{\rm h}01^{\rm m}51\fs6$ $+69\degr35\arcmin27\arcsec$ & $19.6$ & $0.847$ & H${\beta}$  & $124$ & $900/875$ \\
		\hline
	\end{tabular}
	\label{T:sample}
\end{table*}

\begin{figure*}[]
	\includegraphics[width=0.9\linewidth]{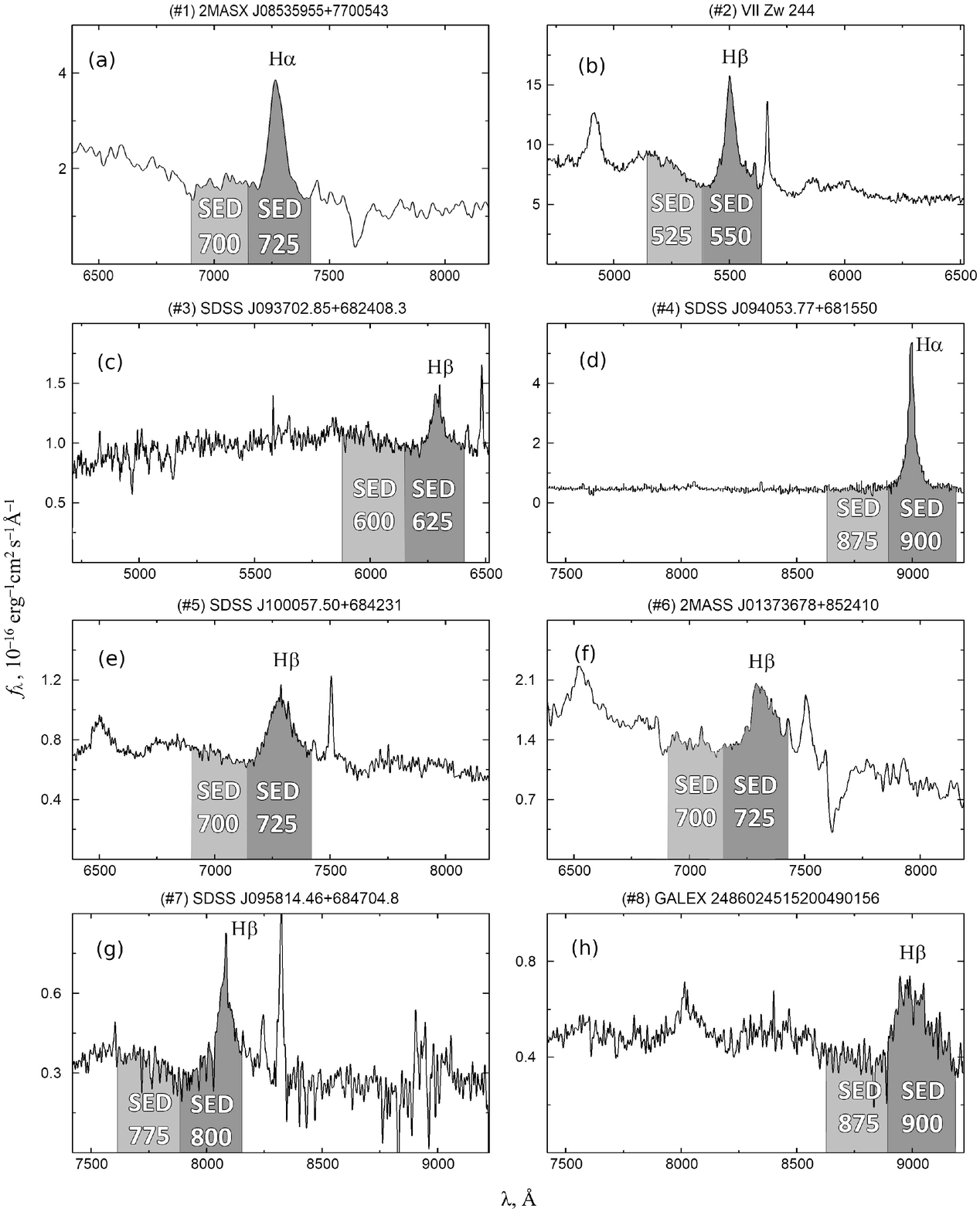}
	\caption{SED bands overplotted on the spectra of the studied AGN.}
	\label{F:filters}
\end{figure*}

Each object is observed in two filters: one corresponds to the
region of the broad emission line H$_{\beta(\alpha)}$, the other
corresponds to the continuum close to the line. Thus, it is
possible to take into account the contribution of the variable
continuum to the observed total flux of the emission line. Thereby
we increase the contrast of the delay of one light curve relative
to another for the cross-correlation analysis. The experiment uses
medium-band interference filters SED\footnote{Edmund Optics,
	https://www.edmundoptics.com} with a 250~\AA\ bandwidth,
overlapping the 5000--9000~\AA\ range also with 250~\AA-step. For
most of the selected objects, a set of filters is used to the
H$_{\beta}$ line and the continuum near it. However, for two
sample objects, \#1 and \#4, the line H$_{\beta}$ fell on the
boundary of neighboring filters, so broad H$_{\alpha}$ line was
chosen instead.  The selection of the filters with their bandwidth
is illustrated in Fig.~{\ref{F:filters}}.  The spectra are taken
from articles \cite{spec,spec1,spec2}.

From the known relations $R_{\rm BLR} - L$ for the H$_{\beta}$
line the expected delays $\tau$ were calculated for the sample
(see Table.~{\ref{T:sample}}). For objects with redshifts up to
0.5---\#\#1,3--5---the flux $L_{\lambda}$ at 5100~\AA\ measured in
the range 4400--5850~\AA\ was calculated. Then from the relation
$R_{\rm BLR}-L_{\lambda}$ (5100~\AA), where $R_{\rm BLR}$ is the
size of the BLR region in the line H$_{\beta}$ \cite{RL5100}:
$$
\log(R_{\rm BLR}) = -21.3^{+2.9}_{-2.8} + 0.519^{+0.063}_{-0.066}
\log(\lambda L_{\lambda}),
$$
where $ L_{\lambda}=L_{\lambda}$ (5100~\AA) is a flux at 5100~\AA.

In the case of $z > 0.5$, as well as for the object \#2, which
spectral data are available only in a small wavelength range
(3500--7000\AA), the $L_{\lambda}$~(5100~\AA) range goes beyond
the available optical spectra. In this regard, for objects
\#\#2,6--8 the calibration in the line
$L_{\lambda}(\text{H}_{\beta})$ \cite{Greene10} was used:
$$
\log(R_{10}) = 0.85 \pm 0.05 + (0.53\pm0.04)
\log(L_{43}(\text{H}_{\beta})),
$$
where $R_{10}=R_{\rm BLR}/10$ lt days is the size of the BLR
region, normalized to the 10~lt days,\linebreak
\mbox{$L_{43}(\text{H}_{\beta}) =
	L_{\lambda}(\text{H}_{\beta})/10^{43}$~erg~s$^{-1}$} is the
luminosity in the H$_{\beta}$  line normalized to
$10^{43}$~erg~s$^{-1}$. In Table \ref{T:sample} the expected
delays $\tau$ are given with an accuracy of 10\%.

\begin{figure}[]
	\includegraphics[width=0.9\linewidth]{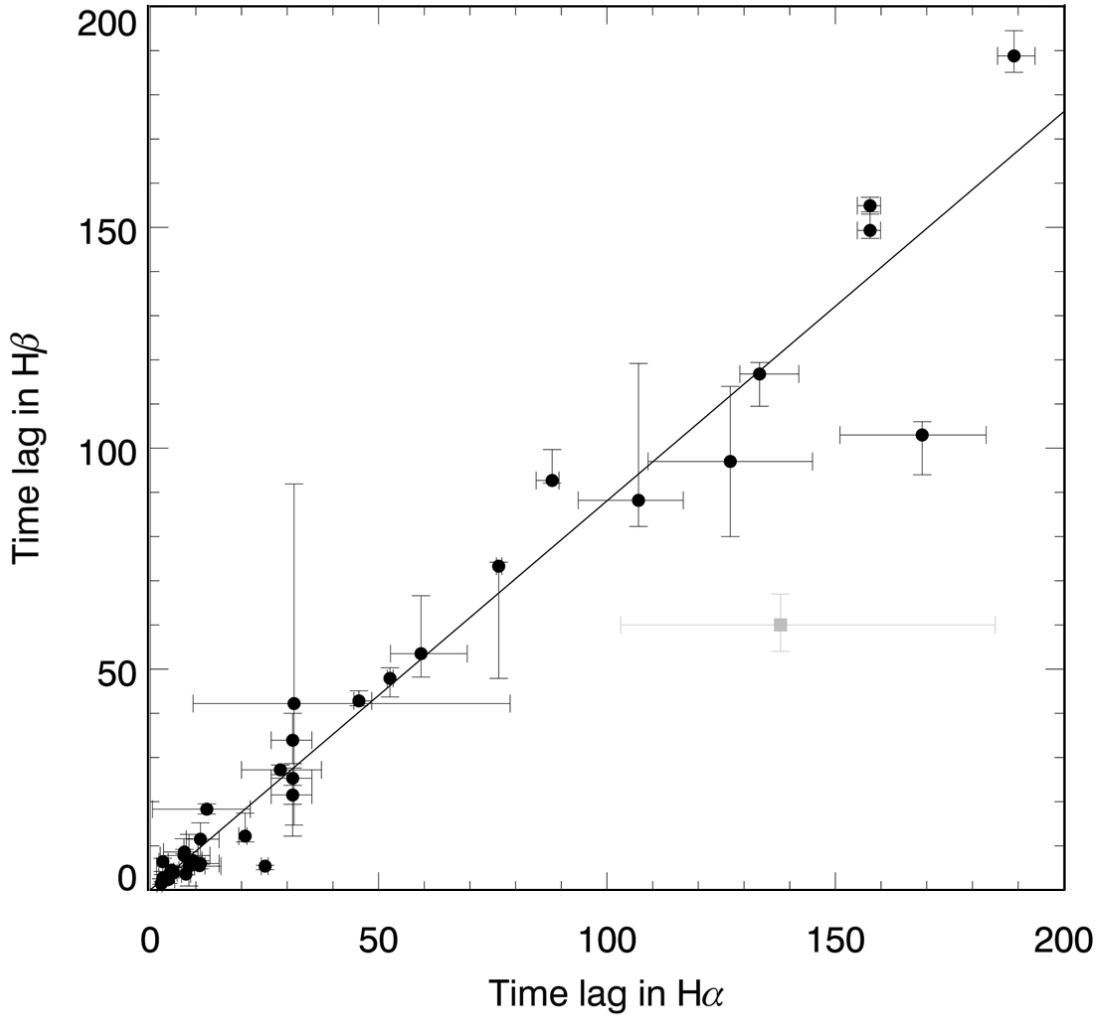}
	\caption{Comparison of the observed time lag in H$_{\beta}$ line
		relative to the time lag in H$_{\alpha}$ line according to the
		catalog \cite{BentzKatz15} (black) and the monitoring \cite{Af15}
		(gray).} \label{F:Hab}
\end{figure}

It is known that the matter in BLR is stratified
\cite{Peterson94, Baldwin}, and the region
emitting in H$_{\alpha}$ is bigger than the region emitting in
H$_{\beta}$. However, the calibration relation
$L_{\lambda}(\text{H}_{\alpha})$ is unpopular since for many AGN
the narrow line N\,II (6583~\AA) belonging to the emission of
narrow-line region clouds (NLR) makes a large contribution to the
H$_{\alpha}$ flux. To estimate the possible difference for the
delay of variation in H$_{\alpha}$ for objects \#1 and \#4, we
used the catalog data \cite{BentzKatz15} for 29 AGN
for delays known in both H$_{\alpha}$ and H$_{\beta}$ lines. Also,
we used data on Sy\,1 3C 390.3 obtained from spectropolarimetric
monitoring on the 6-m BTA telescope \cite{Af15}. A
comparison of the observed lag in the lines is shown in Fig.
\ref{F:Hab}. The slope of the line is equal to $k =
\tau ({\rm H}\beta) / \tau ({\rm H}\alpha) = 0.88 \pm 0.03$. Thus,
the H$_{\alpha}$ lag for \#1 and \#4 coincides with the expected
by H$_{\beta}$ within 10\%.

\section{Observations}
\label{S:obs}

\subsection{Instruments}

Since February 2018, observations of the AGN sample have been
carried out monthly on grey and bright nights at Zeiss-1000
telescope of the SAO RAS using MaNGaL (MApper of Narrow GAlaxy
Lines \cite{Pulkovo}) and MMPP (Multi-Mode Photometer-Polarimeter)
\cite{mmpp} devices in photometric mode with 10 medium-band
interference SED filters. The size of the field of view was
8.$'7\times 8.'7$ for MaNGaL and 7.$'2\times 7.'$2 for MMPP.

Three different detectors were used during the observations: Andor
CCD iKon-M 934 (1024$\times$1024~px), Andor Neo sCMOS
(2560$\times$2160~px), and Raptor Photonics Eagle V CCD
(2048$\times$2048~px). The quantum efficiency of these detectors
in the needed bands is shown in Table~\ref{T:qe}. Water cooling
was used for all three cases to minimize noise.

\begin{table}[]
	\caption{Quantum efficiency of detectors in the studied
		photometric bands}
	\medskip
	\begin{tabular}{l|c|c|c|c|c}
		\hline
		\multicolumn{1}{c|}{\multirow{2}{*}{Detector}}  &   \multicolumn{5}{c}{Quantum efficiency, \%} \\
		\cline{2-6}
		& 5500~\AA\ & 6000~\AA\ & 7000~\AA\ & 8000~\AA\ & 9000~\AA\  \\
		\hline
		Andor iKon-M 934    & 95 & 96 & 91 & 77 & 47 \\
		Andor Neo sCMOS     & 54 & 56 & 49 & 31 & 14 \\
		Eagle V CCD         & 92 & 95 & 89 & 75 & 50 \\
		\hline
	\end{tabular}
	\label{T:qe}
\end{table}

\subsection{Photometric Standards}
Observations of the sample were alternated with observations of
spectrophotometric standard stars from the paper \cite{Oke90}. The
standards were observed before and/or after obtaining a frame with
the object field in the same filter and as close as possible to
the zenith distance. This method of observations makes it possible
to determine the relation between the instrumental units and the
absolute ones beyond the atmosphere and, consequently, to bind the
flux of the selected stars in the field of the object to the
absolute magnitudes to create a network of the local standards.

We have used a system of AB-magnitudes. This system is defined so
that for a monochromatic flux $f_{\nu}$ measured in
erg~s$^{-1}$~cm$^{-2}$~Hz$^{-1}$:
$$m_{AB} = - 2.5 \cdot \log(f_{\nu}) - 48.60. $$
Since the transmittance of SED filters is measured in laboratory,
we denote it as a function $filter(\nu)$ and convolute with a
spectral energy distribution of the photometric standard $f_{\nu}$
to determine its extra-atmospheric AB-value according to the
formula:
$$ m_{AB} = - 2.5 \cdot \log\Bigg[\frac{\int f_{\nu} \cdot filter(\nu) \cdot d\nu}{\int filter(\nu) \cdot d\nu}\Bigg] - 48.60, $$

where the flux of the standard is $f_{\nu}$ also in
erg~s$^{-1}$~cm$^{-2}$~Hz$^{-1}$.

In Table \ref{T:stdoke} the calculated AB-magnitudes of the
observed standards for the used SED filters are given.

\begin{table*}[]
	\caption{The photometric fluxes of the standard stars in
		AB-magnitudes in the SED bands}
	\medskip
	\begin{tabular}{l|c|c|c|c|c|c|c|c|c|c}
		\hline
		\multicolumn{1}{c|}{Stars-}  & \multicolumn{10}{c}{$m_{\rm AB}$, mag}  \\
		\cline{2-11}
		\multicolumn{1}{c|}{standards} & SED\,525 & SED\,550 & SED\,600 & SED\,625 & SED\,700 & SED\,725 & SED\,775& SED\,800 & SED\,875 & SED\,900  \\
		\hline
		G\,$193-74$        & 15.63  & 15.61 & 15.58 & 15.58 & 15.58 & 15.59 & 15.61 & 15.62 & 15.68 & 15.73 \\
		BD\,$+75\degr325$  & 9.47   & 9.56  & 9.73  & 9.81  & 10.03 & 10.10 & 10.22 & 10.28 & 10.46 & 10.51 \\
		Feige\,34          & 11.09  & 11.17 & 11.35 & 11.43 & 11.63 & 11.69 & 11.79 & 11.84 & 11.90 & 12.05 \\
		BD\,$+33\degr2642$ & 10.71  & 10.78 & 10.91 & 10.97 & 11.13 & 11.18 & 11.28 & 11.33 & 11.43 & 11.45 \\
		BD\,$+28\degr4211$ & 10.43  & 10.52 & 10.70 & 10.78 & 10.99 & 11.05 & 11.18 & 11.24 & 11.42 & 11.50 \\
		BD\,$+25\degr4655$ & 9.60   & 9.69  & 9.87  & 9.95  & 10.16 & 10.22 & 10.35 & 10.41 & 10.59 & 10.68 \\
		Feige\,110         & 11.76  & 11.85 & 12.03 & 12.10 & 12.32 & 12.39 & 12.50 & 12.57 & 12.74 & 12.79 \\
		\hline
	\end{tabular}
	\label{T:stdoke}
\end{table*}

For the observational data reduction and subsequent measurements,
the IDL
software\footnote{https://www.harrisgeospatial.com/Software-Technology/IDL}
was used. During each observational night, we received calibration
images (flat frames for each filter at the twilight sky moving the
telescope , bias/dark) to correct data for additive and
multiplicative errors. Photometric standards were also observed at
different zenithal distances to control the extinction coefficient
within the night. To account the light absorption in the
atmosphere, the air masses were calculated according to
\cite{Hardie}:

\begin{equation}\nonumber
\begin{array}{rcl}
M =& \sec(z)&\!\!-0.0018167 [\sec(z) - 1]\\[-5pt]
&         &\!\!-0.002875 [\sec(z) - 1]^{2}\\[-5pt]
&         &\!\!-0.0008083 [\sec(z) - 1]^{3}.
\end{array}
\end{equation}

The method of aperture photometry was used to determine the flux
of objects. Therefore, to correctly account the sky background,
the traces of cosmic rays fell close to the object were removed
from images.

There is a misconception that shooting a sufficiently large number
of frames and summing them it is possible to improve the
signal/noise ratio S/N. The criteria for the correct evaluation of
the S/N ratio are given in the paper \cite{Afanasieva2016}. Since
the image processing has to work with random flux values it is
necessary to correctly determine the estimates. So, each frame is
processed independently, and statistical evaluation is made by
averaging the random value by robust methods giving its unbiased
estimate.

\section{RESULTS}
\label{S:dis}

\subsection{Local Standards}

To measure the absolute AGN variability we have selected the
candidates for local standard stars in the fields of each object.
Over a long period, their brightness must remain constant, plus it
should be comparable to the AGN magnitude to avoid overexposure of
the signal and hence the effects of deviation from linearity on
the detector at long exposures. The use of local comparison stars
for differential photometry significantly increases the accuracy
of measurements of the studied AGN flux, and also allows one to
observe at grey and bright nights and under unstable atmospheric
transparency.

As a result of the first year of monitoring, a network of
comparison stars was formed for photometric binding of AGN under
unstable atmosphere to obtain calibrated light curves in emission
line and continuum. Photometric errors on average do not exceed
$\sigma$ = 0.02 mag.

\begin{table*}[]
	\caption{Comparison stars for the sample objects: (1) coordinates
		for the J2000 epoch; (2) AB-magnitudes of the comparison star in
		the filter corresponding to the observed range of the object
		continuum; (3) AB-magnitudes of the comparison star in the filter
		corresponding to the broad emission line of the object}
	\label{T:compstars}
	\begin{tabular}{c|ccc||c|ccc}
		\hline
		& RA Dec (J2000) & Continuum & Line & & RA Dec (J2000) & Continuum & Line \\
		\hline
		(\#) & (1) & (2) & (3) & (\#) & (1) & (2) & (3)  \\
		\hline
		\#1 & & SED700 & SED725 &  \#5 & & SED700 & SED725 \\
		1-1 & 08$^{\rm h}$54$^{\rm m}$16\fs3 $+76\degr59\arcmin44\arcsec$ & 15.00 $\pm$ 0.01 &  14.95 $\pm$ 0.01 &  5-1 & 10$^{\rm h}$00$^{\rm m}$55\fs4 $+68\degr41\arcmin01\arcsec$ & 16.26 $\pm$ 0.01 &  16.23 $\pm$ 0.01 \\
		1-2 & 08$^{\rm h}$53$^{\rm m}$48\fs5 $+76\degr59\arcmin27\arcsec$ & 15.16 $\pm$ 0.01 &  15.13 $\pm$ 0.01 &   5-2 & 10$^{\rm h}$00$^{\rm m}$50\fs0 $+68\degr40\arcmin32\arcsec$ & 15.70 $\pm$ 0.01 &  15.70 $\pm$ 0.01   \\
		\hline
		\#2 & & SED525 & SED550 &  \#6 & & SED700 & SED725  \\
		2-1 &  08$^{\rm h}$44$^{\rm m}$32\fs0 $+76\degr53\arcmin49\arcsec$ & 12.54 $\pm$ 0.01 &  12.47 $\pm$ 0.01 &  6-1 & 01$^{\rm h}$37$^{\rm m}$15\fs5 $+85\degr22\arcmin28\arcsec$ & 14.82 $\pm$ 0.01 &  14.84 $\pm$ 0.01  \\
		2-2 & 08$^{\rm h}$45$^{\rm m}$22\fs4 $+76\degr50\arcmin12\arcsec$ & 14.03 $\pm$ 0.01 &  13.96 $\pm$ 0.01 &  6-2 & 01$^{\rm h}$36$^{\rm m}$44\fs1 $+85\degr23\arcmin31\arcsec$ & 15.40 $\pm$ 0.01 &  15.43 $\pm$ 0.01 \\
		\hline
		\#3 & & SED600 & SED625 &   \#7 & & SED775 & SED800 \\
		3-1 & 09$^{\rm h}$36$^{\rm m}$44\fs7 $+68\degr25\arcmin46\arcsec$ & 13.73 $\pm$ 0.01 &  13.72 $\pm$ 0.01 &  7-1 & 09$^{\rm h}$58$^{\rm m}$21\fs7 $+68\degr45\arcmin58\arcsec$ & 15.48 $\pm$ 0.01 &  15.45 $\pm$ 0.01    \\
		3-2 & 09$^{\rm h}$36$^{\rm m}$54\fs6 $+68\degr24\arcmin39\arcsec$ & 16.63 $\pm$ 0.07 &  16.60 $\pm$ 0.06 &  7-2 & 09$^{\rm h}$58$^{\rm m}$45\fs4 $+68\degr45\arcmin09\arcsec$ & 16.93 $\pm$ 0.01 &  16.84 $\pm$ 0.01   \\
		\hline
		\#4 & & SED875 & SED900 &  \#8 & & SED875 & SED900 \\
		4-1 & 09$^{\rm h}$40$^{\rm m}$51\fs8 $+68\degr15\arcmin10\arcsec$ & 15.57 $\pm$ 0.02 &  15.54 $\pm$ 0.02 &  8-1 & 10$^{\rm h}$01$^{\rm m}$56\fs4 $+69\degr32\arcmin46\arcsec$ & 16.13 $\pm$ 0.02 &  16.16 $\pm$ 0.03  \\
		4-2 & 09$^{\rm h}$41$^{\rm m}$06\fs9 $+68\degr16\arcmin41\arcsec$ & 14.99 $\pm$ 0.02 &  14.95 $\pm$ 0.02 & 8-2 & 10$^{\rm h}$02$^{\rm m}$04\fs6 $+69\degr34\arcmin02\arcsec$ & 17.26 $\pm$ 0.02 &  17.19 $\pm$ 0.04  \\
		\hline
	\end{tabular}
\end{table*}

The results obtained for all comparison stars are summarized in
Table \ref{T:compstars}.

\subsection{Preliminary Result}

\begin{figure}[]
	\includegraphics[width=0.5\textwidth]{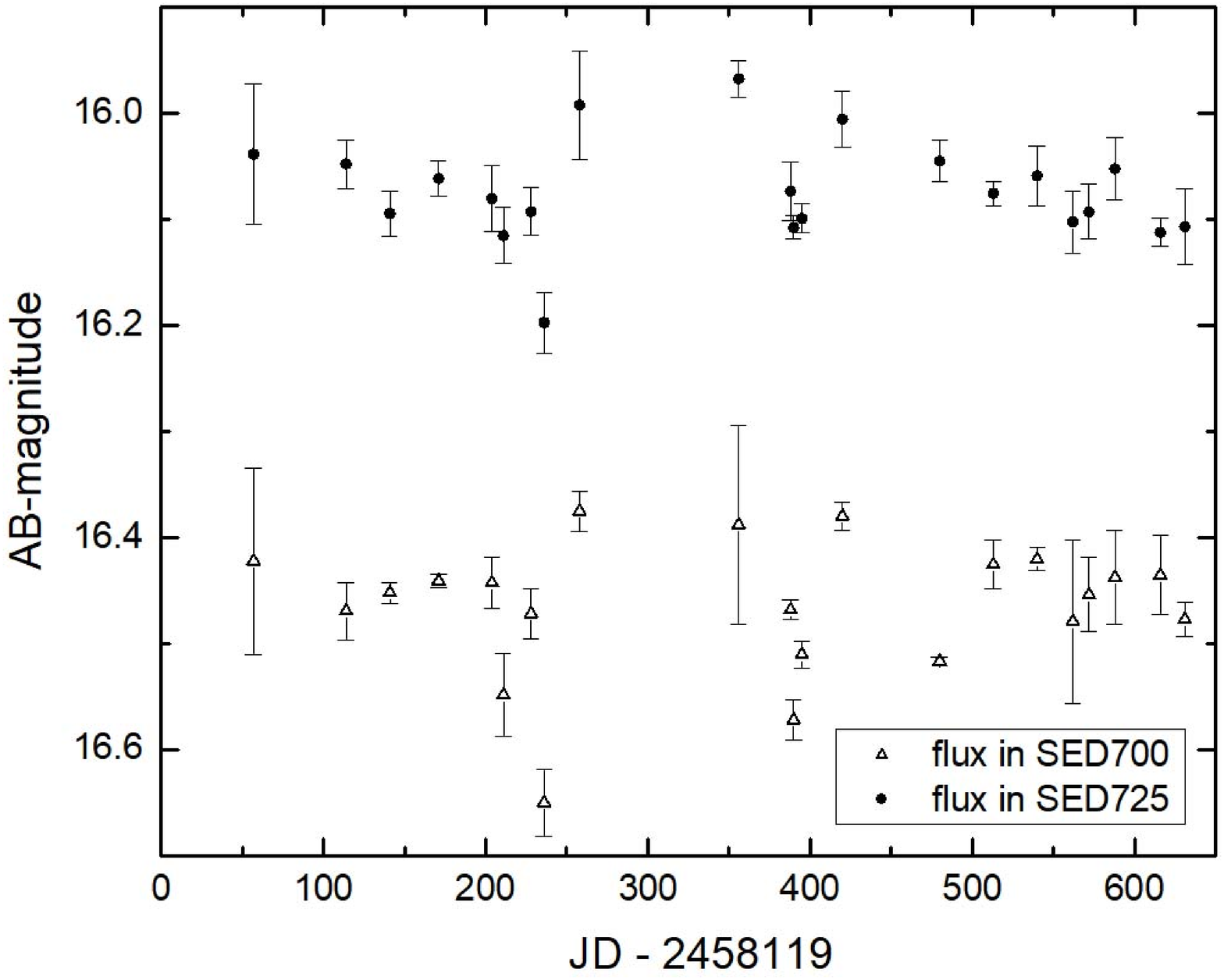}
	\caption{The light curves of AGN \#1, obtained in the SED filters
		corresponding to the flux in the line and in the continuum near.
		The Julian dates starts on January 1, 2018.}
	\label{F:LC}
\end{figure}


The measurements of the studied AGN fluxes were carried out
relative to the most stable reference stars assumed to be the
local standards. The light curves in the continuum and the line of
one of the most frequently observed AGN \#1 (2MASX
J08535955+7700543) are shown in Fig.~{\ref{F:LC}}.

After subtracting the continuum flux from the total flux in the
line, there is a short-term variability at the level of 0.2
AB-magnitudes of the both fluxes, and the character of the
variability is repeated. The observed amplitude exceeds the
average error of AGN radiation measurement: the differential
photometry method provides an accuracy of $\sim$0.03 mag.

To estimate the time delay between the two light curves of the
object \#1 of the sample, the classical cross-correlation method
ICCF was used as well as the method using the code {\tt JAVELIN}
\cite{Zu16,jav}. The results are presented
in Fig.~\ref{CCF}.

\subsubsection{Classical Cross-Correlation Method}

The solid curve in Fig.~\ref{CCF} denotes the interpolated
cross-correlation function (ICCF). Fitting the Gaussian to the
most powerful ICCF peak gives us an estimate of the time delay
$\tau (ICCF)=32.2\pm10.6$~days. In this estimate, we use the
half-width of the Gaussian interpolation as the measurement error.
Note that to obtain a contrast peak, it is also necessary to
subtract the contribution of the continuum to the total flux in
the emission line.

\subsubsection{JAVELIN}

\begin{figure}[]
	\includegraphics[width=0.95\linewidth]{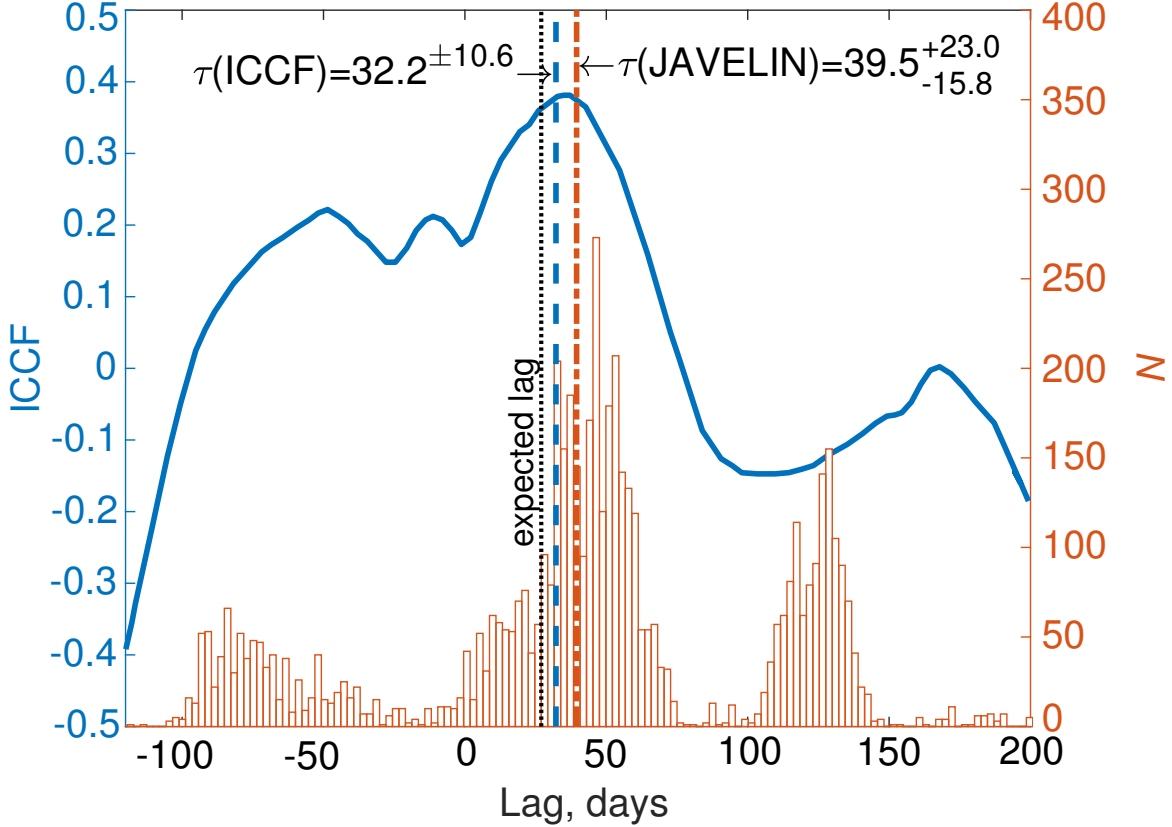}
	\caption{Cross-correlation analysis of the time delay between the
		continuum flux and the flux in the emission line for
		2MASS\,J08535955+7700543 (\#1) using two methods: classical ICCF
		(solid interpolated curve) and JAVELIN (histogram obtained using
		{\tt JAVELIN}).    Corresponding time delay estimates marked by
		dashed ($\tau(ICCF)$) and dash-dotted ($\tau({\rm JAVELIN})$)
		vertical lines. The dotted line shows the expected value of $\tau$
		from calibrations.}
	\label{CCF}
\end{figure}

Fig.~\ref{CCF}  shows the JAVELIN method as a histogram obtained
using {\tt JAVELIN} (Just Another Vehicle for Estimating Lags in
Nuclei) code implemented in the {\tt python} programming language.
We describe briefly the content of the procedure for determining
$\tau$ using this method. The first step is to build a continuum
model using the DRW (dumped random walk) method. As a result, we
have posterior distributions of two DRW parameters of continuum
variability---amplitude and time scale calculated on the basis of
MCMC sampling (Markov chain Monte Carlo)\footnote{MCMC---an
	algorithm to generate a sample from a posteriori probability
	distribution and compute integrals by Monte Carlo method. The
	sequence of values obtained from a reversible Markov chain whose
	stable distribution is the target posterior distribution.}. The
second step is to interpolate the light curve of the continuum
based on the parameters defined in the first step and then offset,
smooth, and scale it to compare with the observed line light
curve. After another run of the MCMC algorithm, the {\tt JAVELIN}
package determines the desired posterior time delay distribution
between the light curves. As a result, we got the value $\tau({\rm
	JAVELIN})=39.5_{-15.8}^{+23.0}$ days. The estimate itself
corresponds to the median value of the most powerful peak, located
in the range from $ -20$ to 80 days in Fig.~\ref{CCF}. The lower
and upper estimates of the time delay correspond to the limits of
the highest density interval of the posterior distribution, which
are calculated using {\tt JAVELIN}.

\subsubsection{General Comment}

So, to illustrate the technique efficiency, we used AGN \#1 light
curves and revealed estimates of the time delay
$\tau(ICCF)=32.2^{\pm10.6}$~days and \mbox{$\tau ({\rm
		JAVELIN})=39.5_{-15.8}^{+23.0}$}~days. Within the limits of
accuracy, our estimates are in good agreement with each other and
with the expected time delay $\tau\approx27$ days from the
calibration relations. Despite the fact that cross-correlation
peaks are confidently detected, we assume a continuation of the
accumulation of observational data for the light curves to refine
the result of AGN~\#1 cross-correlation analysis. A direct
comparison of the time delay error values $\Delta\tau$ for ICCF
and JAVELIN methods is inappropriate and requires additional
research within this project.  Both methods work well even in the
presence of systematic errors \cite{Yu19}.

It should be noted that the measurement error of the delay
$\Delta\tau$ is closely connected with the sampling period of the
light curves $t_{cad}$, i.e., the time between sets of
observations \cite{Zu16}. Over the past year, the average period
of $t_{cad}$ was $\sim$20-25 days. To specify the value of the
delay $\tau$ it is necessary to increase the number of sets of
observations, thereby reducing the sampling, which is especially
important for active nuclei with the expected delay of the
radiation with the lags of the order of several tens of days, for
example, \#1 and \#2.

\section{Conclusions}
\label{S:con}

Within this work, the following results were obtained.

\begin{list}{}{
		\setlength\leftmargin{2mm} \setlength\topsep{2mm}
		\setlength\parsep{0mm} \setlength\itemsep{2mm} }
	
	\item 1. The observations by the photometric reverberation mapping
	method are adapted for telescopes of 1-meter class and are
	independent of the device used.
	
	\item 2. For each of the studied active nuclei in the range of
	redshifts $0.1 < z < 0.8$, a network of secondary standards was
	determined, which allows further use of the differential
	photometry method. The photometric accuracy is on average 0.03
	mag, which is an order of magnitude greater than the expected
	amplitude of the AGN variability.
	
	\item 3. Preliminary results of the object
	2MASX\,J08535955+7700543 (\#1) reverberation mapping are shown in
	Fig.~\ref{F:LC}. It is seen that the observed object is variable,
	and the used method is stable. Applying the classical
	cross-correlation function and {\tt JAVELIN} gave estimates of the
	time delay $\tau(ICCF)=32.2^{\pm10.6}$ days and\linebreak
	\mbox{$\tau({\rm JAVELIN})=39.5_{-15.8}^{+23.0}$} days that are
	consistent with each other and within the accuracy of the existing
	calibration relations.
\end{list}

\section{Acknowledgments}
The authors are sincerely grateful to the reviewer for fruitful
comments, which contributed to the improvement to the article.

The authors thank V. L. Afanasiev for useful discussions and
comments.

\section*{FUNDING}
The work is executed at support of RFBR grant 18-32-00826.
Observations on the telescopes of SAO RAS are carried out with the
support of the Ministry of science of the Russian Federation.

\section*{CONFLICT OF INTERESTS}

The authors declare no conflict of interest regarding this paper.

\bibliographystyle{AstroBull}
\bibliography{sample_eng}

%
%
%
%
%



\onecolumngrid
\clearpage

\selectlanguage{english}
\begin{center}
\end{center}
\begin{center}
\end{center}
\begin{center}
\end{center}
\begin{center}

\end{center}

\end{document}